\documentclass[preprint,aps]{revtex4}

\usepackage{graphicx}
\usepackage{dcolumn}

\begin{document}

\preprint{Lebed-Naughton}

\title{Interference Commensurate Oscillations in Q1D Conductors}

\author{A.G. Lebed$^{1,2}$ and M.J. Naughton$^1$}

 \email{lebed@bc.edu, naughton@bc.edu}

 \affiliation{$^1$Department of Physics, Boston College, Chestnut Hill, 
MA 02467, USA}

\affiliation{$^2$Landau Institute for Theoretical Physics,
2 Kosygina Street, 117334 Moscow, Russia}

\date{February 24 2003, Submitted to Physical Review Letters}

\begin{abstract}
We suggest an analytical theory to describe 
angular magnetic oscillations recently discovered in 
quasi-one-dimensional conductor (TMTSF)$_2$PF$_6$
[see Phys. Rev. B \underline{57}, 7423 (1998)] and define the positions 
of the oscillation minima. 
The origin of these oscillations is related to interference effects 
resulting from an interplay of quasi-periodic and periodic ("commensurate") electron trajectories in an inclined 
magnetic field. 
We reproduce via calculations existing experimental data and predict 
some novel effects. 
\end{abstract} 


\maketitle

\pagebreak

Quasi-one-dimensional (Q1D) organic conductors (TMTSF)$_2$X 
(X = PF$_6$, ClO$_4$, etc.) demonstrate a variety of unique 
properties in a magnetic field in their superconducting [1-3], 
field-induced spin-density-wave  [1-3], and metallic 
[4-22,1-3] phases. 
In the metallic phase [4-22,1-3], the
Q1D electron spectra [1-3] of these compounds,

\begin{equation}
\epsilon^\pm ({\bf p}) = \pm v_F \ (p_x \mp p_F) + 2 t_b
 \cos(p_y b^*) + 2 t_c \cos(p_z c^*) \ , v_F p_F \gg t_b \gg t_c \ ,
\end{equation}
are characterized by two open sheets of Fermi surface (FS).
Therefore, traditional magnetic oscillations (related to Landau 
level quantization) [23] cannot exist in these materials. 
[Here $+(-)$ stands for the right (left) sheet of the 
Fermi surface, $v_F$ and $p_F$ are the Fermi velocity and Fermi 
momentum, respectively; $t_b$ and $t_c$ are the overlapping
integrals between electron wave functions; $\hbar \equiv 1$].

Surprisingly, the metallic phases of (TMTSF)$_2$X materials 
exhibit a number of unconventional magnetic oscillations 
related to an open Q1D FS (1). 
Among them are "magic angles" (MA) 
[4-9,1-3], the first angular oscillations with a clear Fermi-liquid (FL) physical meaning - Danner-Kang-Chaikin (DKC) oscillations [10], 
the "third angular effect" (TAE) [11-19], and rich angular oscillations 
recently discovered by Lee and Naughton [12,15,16] in (TMTSF)$_2$PF$_6$ 
and by Yoshino et al. [17] in (DMET)$_2$I$_3$. 
We call the latter "interference commensurate" (IC) oscillations 
which reflects their physical meaning revealed in this
Letter.  
Note that, despite all these observations of different magnetic 
oscillations [4-19], a question on the existence or not of FL behavior in Q1D metallic phases of (TMTSF)$_2$X conductors
is still controversial [3,9,20-22].

Numerical solutions [15,16,18,19] of the 
Boltzmann kinetic equation for Q1D metal (1) in an inclined 
magnetic field,
\begin{equation}
{\bf H} = H ( \cos \theta \cos \phi, - \cos \theta \sin \phi, 
\sin \theta) \ ,
\end{equation}
have given a hint on the FL nature of the IC 
oscillations.
Nevertheless, due to very complex behavior of these oscillations, 
their physical meaning has not been revealed and 
their properties have not been described in detail.
In particular, a hypothesis [15] that minima of resistivity 
$\rho_{zz}({\bf H})$ correspond to some "commensurate" directions of 
a magnetic field,
\begin{equation}
\sin \phi = N \biggl( \frac{b^*}{c^*} \biggl) \tan \theta  \ ,
\end{equation}
(where N is an integer) has not been theoretically proven.

The goals of our Letter are: 1) to derive an analytical expression 
for $\rho_{zz}({\bf H})$ and to define
the positions of its minima; 2) to reveal a quantum interference 
nature of the IC oscillations, 3) to compare our results with experiment [15], and 
4) to predict some novel qualitative effects.
In particular, we demonstrate that the origin 
of the IC oscillations is related to special "commensurate" 
electron trajectories in a magnetic field (different from the MA trajectories [4]), 
where an average electron velocity along the ${\bf z}({\bf c^*})$-axis is non-zero, 
$< v_z(t)>_{t} \neq 0$ (see Fig.1).
Note that an importance of this condition for low-dimensional conductors 
was pointed out by DKC [10], Osada et al. [24], 
and by Kartsovnik and Yakovenko et al. [25] in different context.

Below, we consider the most general electron trajectories, 
characterized by two angles, $\theta$ and $\phi$ [see Eq. (2)]. 
We show that they correspond to two novel types of angular magnetic
 oscillations: "commensurate" oscillations and some "generalized DKC" oscillations.
In particular, we demonstrate that $\rho_{zz}({\bf H})$ in this case
is defined by quantum interference effects related to periodic
and quasi-periodic electron motion along the open Q1D FS (1).
For small values of $\theta$ [see Eq.(2)], these effects occur 
between some narrow areas on the Q1D FS (1) 
[i.e., "effective stripes" (ES) parallel to ${\bf p}_z$-axis ], with 
their positions being dependent on the magnetic field orientation 
(see Fig.1c).
These unique features demonstrate an unusual physical meaning of 
the IC oscillations.
We also show that the "commensurate" directions of a magnetic
field (3) indeed correspond to minima in $\rho_{zz}({\bf H})$ 
at large enough $ \theta $, whereas at smaller $ \theta $ 
($|\theta | \leq 5^o$ in our case) $\rho_{zz}({\bf H})$ minima 
occur for only even or only odd values of N in Eq. (3), depending on
the particular value of $\theta$.

Let us discuss the physical meaning of the 
IC oscillations by analyzing quasi-classical electron trajectories 
in an inclined magnetic field (2).
As usual, the trajectories are solutions
of the equations of motion [23]:

\begin{equation}
d {\bf p} / dt =  (e / c)  \ [{\bf v}({\bf p}) \times {\bf H}] \ , 
\ \ \ \ \ \ \ \ \ 
{\bf v}({\bf p}) = d \epsilon ( {\bf p} ) / d {\bf p}
\ .
\end{equation}
For Q1D electron spectrum (1), Eqs.(4) can be rewritten as 
follows [26]:

\begin{equation}
d(p_y b^*)/dt =  \omega_b (\theta) \ , \ \ \ \ 
d(p_z c^*)/ dt = \omega_c (\theta , \phi) - 
\omega^*_c (\theta , \phi)  \sin (p_y b^*) \ ,
\end{equation}
where
\begin{eqnarray}
&&\omega_b (\theta) = e v_F H b^* \sin \theta / c \ ,
\ \ \ \ \omega_c ( \theta , \phi ) = 
e v_F H c^* \cos \theta \sin \phi / c \ ,
\nonumber\\
&&\omega^*_c ( \theta , \phi ) = 
(v^o_y / v_F) (e v_F H c^*/c)
 \cos \theta \cos \phi , \ \ \ \ v^o_y = 2 t_b b^* .
\end{eqnarray}
From Eqs.(5), one defines electron trajectories in a reciprocal plane 
$(p_y , p_z)$:
\begin{equation}
p_z c^* = p^o_z c^* + \frac{\omega_c (\theta , \phi)}
{\omega_b (\theta )} (p_y - p^o_y) b^* + 
\frac{\omega^*_c (\theta , \phi)}{\omega_b (\theta )} 
[ \cos(p_y  b^*) - \cos(p^o_y  b^*)] \ ,
\end{equation}
with a velocity component along the ${\bf z}$-axis being
\begin{eqnarray}
v_z(p_y) = - 2 t_c c^*
\sin \biggl( 
&&p^o_z c^* + \frac{\omega_c (\theta , \phi)}
{\omega_b (\theta )} (p_y - p^o_y) b^* 
\nonumber\\ 
&&+ \frac{\omega^*_c (\theta , \phi)}{\omega_b (\theta)} 
[ \cos(p_y  b^*) - \cos(p^o_y  b^*)]
\biggl)  \ , \ \ \  p_y \sim t .
\end{eqnarray}
Note that electron trajectories in a reciprocal plane
$(p_y , p_z)$ become periodic [27] for "commensurate" 
directions of a magnetic field (3) (see Figs.1a,b).
Therefore, an average velocity over electron path,
 
\begin{equation}
<v_z(t)>_t \sim 
\sin \biggl[ p^o_z c^* - \frac{\omega_c(\theta , \phi)}
{\omega_b (\theta )} p^o_y b^* - 
\frac{\omega^*_c (\theta , \phi)}{\omega_b (\theta )} 
\cos(p^o_y b^*) \biggl] \  
J_{N} \biggl[ \frac{\omega^*_c (\theta , \phi)}
{\omega_b (\theta )} \biggl] \ ,
\end{equation}
(which is zero for all "non-commensurate" trajectories), for 
"commensurate" orbits (3,7), becomes zero only at zero values of 
the N-order Bessel functions:
\begin{equation}
J_N \biggl[ \frac{\omega^*_c (\theta , \phi)}
{\omega_b (\theta )} \biggl] \equiv
J_N \biggl[ \frac{2 t_b c^* \cos \phi}{v_F \tan \theta} \biggl] 
= 0 \ .
\end{equation}

From general metals theory [23,10], it is known that $ <v_z(t)>_t \neq 0$
results in saturating behavior of the conductivity $\sigma_{zz}(\theta , \phi)$
for high magnetic fields, whereas, at $ <v_z(t)>_t = 0$, 
$\sigma_{zz}(\theta , \phi)$ decreases with increasing field.
Therefore, there appear maxima in $\sigma_{zz}(\theta , \phi)$ 
[i.e., minima in $\rho_{zz}(\theta , \phi)$] at "commensurate" angles (3) 
(see Fig.2) which are a novel type of angular magnetic oscillations.
In a similar way, Bessel functions zeros (10) lead to the
appearance  of maxima in $\rho_{zz}({\bf H})$ (see Fig.2) which 
are a generalization of the DKC oscillations corresponding to 
$N=0$ in Eq. (10).
In other words, the IC oscillations are characterized by minima
in $\rho_{zz}({\bf H})$ at "commensurate" angles (3) which are 
modulated by oscillatory Bessel functions (10).

Let us discuss electron motion along the ${\bf z}$-axis in real 
space, $ z(t) = \int^t_0 v_z(t^{'}) d t^{'}$.
Here, contributions to $z(t)$ from different Brillouin zones [see Eq.(8)] 
are in-phase for "commensurate" trajectories (3) (see Fig.1c).
These interference effects correspond to maxima of conductivity $\sigma_{zz}(\theta , \phi)$ [i.e., minima of resistivity 
$\rho_{zz}(\theta , \phi)$] (see Fig. 2).
For $ \omega^*_c(\theta , \phi) / \omega_b(\theta ) \gg 1 $ (i.e., for
small $ \phi $), the integral $z(t) = \int^t_0 v_z(t^{'}) d t^{'}$ can be
evaluated by means of the stationary-phase method and is determined by 
interference effects between some ES located near the points
$p_{y} b^* = \arcsin [\omega_c( \theta , \phi) / \omega^*_c( \theta , \phi )]
+ 2 \pi n$ 
and 
$ p_{y}b^* = \pi - \arcsin [\omega^*_c (\theta , \phi) / 
\omega_c(\theta , \phi) ] + 2 \pi n$ 
on Q1D FS (1) (see Fig.1c), where $n$ is an integer.

To develop an analytical theory of the IC oscillations and to further demonstrate
their quantum interference nature, we make use of the quasi-classical approximation for electron motion [23,28], where a magnetic field is introduced by the Peierls substitution [23,28],  
$ {\bf p} \rightarrow {\bf P} - (e/c) {\bf A}$.
We choose a vector-potential of magnetic field 
(2), in the following form
 
\begin{equation}
{\bf A} = ( 0 , x \sin \theta , x \cos \theta \sin \phi +
y \cos \theta \cos \phi ) \ H \ .
\end{equation}
In this case, electron wave functions in a mixed $(p_y,x)$-representation 
[28], 
$\Psi^{\pm}_{\epsilon }(p_y,x) = \exp(\pm i p_F x) \psi^{\pm}_{\epsilon }  (p_y,x)$,
are solutions of the Schrodinger equations [28],
\begin{equation}
\biggl( 
\mp i v_F \frac{d}{ dx} + 2 t_b \cos \biggl[ p_y b^* - \frac{\omega_b(\theta) x}{v_F} \biggl] 
\biggl) \psi^{\pm}_{\epsilon}  (p_y,x) = 
\epsilon \psi^{\pm}_{\epsilon}    (p_y,x) \ ,
\end{equation}
and can be expressed as
\begin{equation}
\psi^{\pm}_{\epsilon } (p_y, x) = 
\exp \biggl( \pm i \frac{\epsilon}{v_F} x \biggl ) 
 \exp \biggl[ \pm \frac{2 i t_b }{\omega_b(\theta)} 
\biggl( \sin \biggl[ p_y b^* - \frac{\omega_b ( \theta ) x}{ v_F} \biggl] - 
\sin [p_y b^*] \biggl)  
\biggl] \ .
\end{equation}
Note that the ${\bf z}$-component of the quasi-classical velocity operator, 
${\hat {\bf v}}( {\bf p} ) 
= d { \hat \epsilon } ({\bf p}) / d {\bf p}$ [23], 
in a gauge (11) is equal to

\begin{equation}
{\hat v_z} (p_z , x,y) = -2 t_c c^* \sin \biggl[ p_z c^* - 
\frac {\omega_c(\theta , \phi) x}{ v_F} 
- \frac {\omega^*_c(\theta , \phi) y }{ v^0_y } \biggl] \ , 
\ \ \ \ y = i \frac {d}{p_y} \ .
\end{equation} 
It is possible to show that wave functions (13) are eigenfunctions  of velocity operator 
(14) with their eigenvalues being: 
\begin{eqnarray}
{\hat v_z} (p_z , x,y) \psi^{\pm}_{\epsilon } (p_y,x) = -2 t_c c^*
\sin \biggl[&&p_z c^* -  \frac{\omega_c(\theta , \phi) x}{v_F} 
\pm \frac{\omega^*_c(\theta , \phi)}{\omega_b (\theta)} 
\nonumber \\ 
\times &&\biggl( \cos \biggl[ 
p_y b^* - \frac{\omega_b (\theta) x}{ v_F} 
\biggl] 
- \cos [p_y b^*] \biggl) 
\biggl]
\psi^{\pm}_{\epsilon } (p_y,x) \ .
\end{eqnarray}
Since wave functions (13) and their matrix elements (15) 
are known, $\sigma_{zz}(H, \theta, \phi)$ can be evaluated by means of 
the Kubo formalism.
After straightforward but rather routine calculations, we obtain:
\begin{eqnarray}
\sigma_{zz}(H, \theta, \phi) &&\sim 
\int^{ 2 \pi }_{0} \ d(p_y b^*) \
\int^{ \infty }_0 \ d x \  
\exp \biggl(  - \frac{x}{ v_F \tau} \biggl) 
\nonumber \\
&&\times \cos \biggl( -\frac{\omega_c(\theta , \phi) x}{v_F} + 
\frac{\omega^*_c (\theta , \phi)}{\omega_b (\theta)}
\biggl[ \cos \biggl( p_y b^* - \frac{\omega_b(\theta) x}{ v_F} \biggl) - 
\cos(p_y b^*) \biggl] \biggl) \ , 
\end{eqnarray}
where $\tau$ is an electron relaxation time.
Double integral (16) can be rewritten in a simple form:

\begin{equation}
\sigma_{zz}(H, \theta, \phi)  = \sigma_{zz}(0) 
\sum^{+ \infty}_{N= - \infty} 
\biggl( \frac{J^2_N [ \omega^*_c(\theta , \phi) / \omega_b(\theta )]}
{ 1 + \tau^2 [\omega_c (\theta , \phi) - N \omega_b( \theta ) ]^2} 
\biggl) \ ,
\end{equation}
where for a Q1D conductor (1):
 \begin{equation}
\rho_{zz} (H, \theta, \phi) \simeq 1 / \sigma_{zz}(H, \theta, \phi) \ .
\end{equation}

To summarize, Eqs.(17) and (18) provide analytical expressions for
experimentally measured resistivity $\rho_{zz} (H, \theta, \phi)$ 
[15,16].
As is seen from Eq. (17), $\sigma_{zz} (H, \theta , \phi)$
possesses maxima [i.e., $\rho_{zz} (H, \theta , \phi)$) possesses minima]
at $ \omega_c (\theta , \phi) = N \omega_b( \theta ) $ if 
$ \omega_c(\theta , \phi), \omega_b(\theta) \geq 1 / \tau $.
This coincides with the "commensurability" condition (3).
It is important to note that this theory, based on Eqs.(17),(18), predicts no angular 
oscillations [29] at MA directions of a magnetic field [4] (i.e., at $\phi = \pi/2$) 
since $\omega^*_c = 0$ in Eq.(17).
Therefore, the previous interpretation [15-19] of the IC oscillations 
as a simple combination of the MA effects and the DKC oscillations is too oversimplified.   
We also stress that the integration of the quasi-periodic function in Eq.(16)
corresponds to interference effects in matrix elements of velocity operator (15) and, thus, directly demonstrates the interference nature of the
 IC oscillations.
Indeed, two different periods in Eq.(16) become commensurate at 
$\omega_c(\theta, \phi) = N \omega_b(\theta, \phi)$ [i.e., for "commensurate" trajectories (3)] and, therefore, the integral (16) is increased due to these interference effects.

Let us consider Eq.(17) at small enough angles,
$\phi \ll \pi/2$ and $\theta \ll 2 t_b c^*/v_F$.
In this case, one can make use of an asymptotic expression for the Bessel
functions in Eqs.(10),(17): $J_N(2 t_b c^*/v_F \tan \theta) \simeq  
\cos(2 t_b c^*/v_F \tan \theta - \pi/4 -\pi N /2)$.
Therefore, depending on the value of parameter $2t_bc^*/v_F \tan \theta$, the Bessel functions of even orders
are bigger than those of odd orders or vise versa.
At $\omega_c \tau , \omega_b \tau \simeq 1$, this results in the
appearance of minima of $\rho_{zz}(\phi , \theta)$ in Eqs.(17),(18)
only for all even or only for all odd values of the integer $N$.
We call this phenomenon "even-odd" angular resonance (see Fig.2b). 
It is important that Eq.(17) demonstrates also another kind of angular 
oscillations (i.e., "extended DKC" oscillations) related to zeros of the 
Bessel functions (10).
By analyzing Eqs.(16)-(18), it is possible to show that $\rho_{zz}(H, \theta,\phi)$ is characterized by an unusual linear behavior for "non-commensurate" directions of a magnetic field and small $ \theta \ll 2 t_b b^* / v_F $,
\begin{equation}
\rho_{zz} (H, \theta, \phi) \sim |H| ,
\end{equation}
whereas, for commensurate directions (3), $\rho_{zz}(H, \theta,\phi)$ saturates with increasing magnetic field.
The latter effect is different from the linear magnetoresistance predicted in
Ref. [14].

In Fig.2, we compare the experimental data [15] with Eqs. (17),(18)
using the same values of parameters,
$t_a / t_b = 8.5$ and $ \omega_c(\theta = 0,\phi = \pi/2, H = 1 \ T) \ \tau 
= 15 $, for all three theoretical curves.
These curves not only demonstrate qualitative but quantitative agreement 
between theory (17),(18) and experiment [15], in a broad region of magnetic field orientations, $\phi \leq 20^o$ [26].
Note that, for $\theta = 3^o$, $\rho_{zz} (H, \theta , \phi)$ minima appear both theoretically and experimentally only for odd integers $N$ in Eq.(3) [see Fig.(2)] which is an agreement with the "odd" angular resonance effect discussed above.
The agreement is far better than what has been achieved with numerical
solutions of the kinetic equation [15-19], which do not account in full for 
the interference nature of the IC oscillations.
We have initiated more detailed experiments to try to confirm other
effects predicted in the Letter such as "even" angular resonance, "generalized DKC" oscillations (10), and linear magnetoresistance (19).

Let us discuss peculiarities of the interference effects at small angles
$ \theta \ll 2 t_b b^* / v_F $   in a clean limit where 
$\omega_c(\theta , \phi) \tau , \omega_b(\theta) \tau \rightarrow \infty $.
In this case, the integral (16) can be evaluated by means 
of the stationary-phase method and is determined by the close proximity of the 
following two series of "effective points" (EP) in $ (p_y ,x) $-space:  
\begin{eqnarray} 
\omega_b (\theta) z_{n} / v_F  =  \arcsin [ \omega_c (\theta , \phi) / 
\omega^*_c (\theta , \phi) ] + 2 \pi n , \ z = 
x - v_F p_y b^*/ \omega_b( \theta) \ ,
\\  
\omega_b (\theta) z_{n} / v_F = \pi - \arcsin [ \omega_c (\theta , \phi) / 
\omega^*_c (\theta , \phi) ] + 2 \pi n , \ z = 
x - v_F p_y b^*/ \omega_b( \theta) \ ,
\end{eqnarray}
where $n$ is an integer.
In quantum mechanical language, these EP (20),(21) correspond to ES on the Q1D FS (1) discussed above.
The contributions to (16) from the EP (20),(21) are characterized by non-zero 
phase shifts with their values being dependent on magnetic field orientation (2).
Note that these phase shifts are integer values of $2 \pi$ for different 
$n$ in Eqs.(20),(21) only at 
$\omega_c(\theta, \phi) = N \omega_b (\theta , \phi)$.
Therefore, at small angles $\theta$, the IC oscillations (3),(10) can be interpreted in terms of interference effects between these two infinite
 series of electron waves (20),(21). 
It is easy to show that, in our picture, different integers $n$ in Eqs.(20),(21)
 correspond to different Brillouin zones for a quasi-classical electron motion 
picture in the extended Brillouin zone.
 
This work was supported in part by National Science Foundation, grant number
DMR-0076331, the Department of Energy, grant number DE-FG02-02ER63404, and
by the INTAS grants numbers 2001-2212 and 2001-0791.
One of us (AGL) is thankful to N.N. Bagmet and E.V. Brusse for useful discussions.

\pagebreak

\begin{figure}[h]
\includegraphics[width=2.8in,clip]{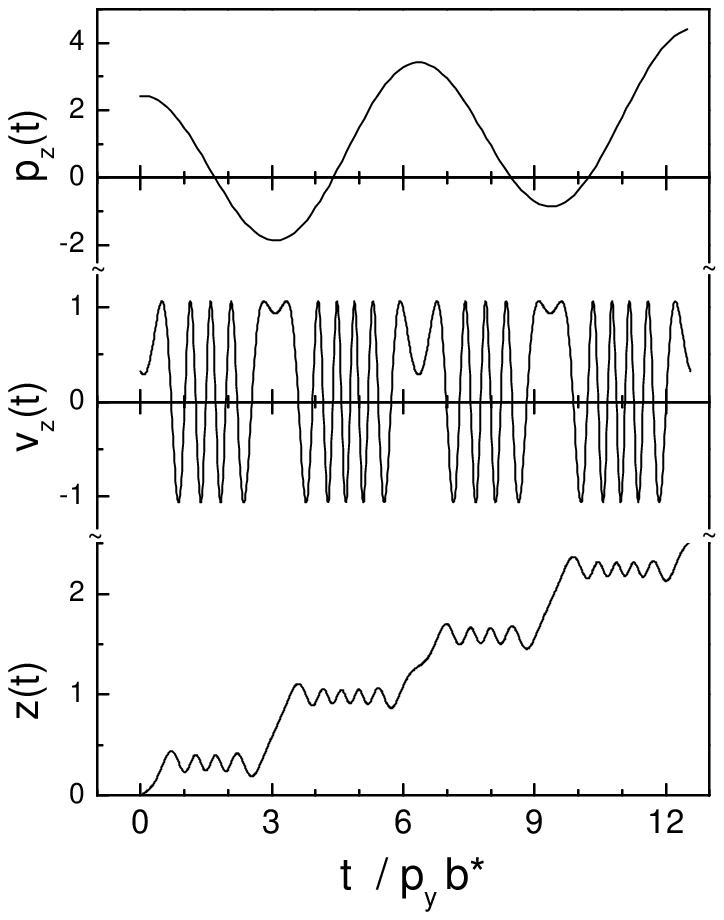}
\caption{ For a "commensurate" electron trajectory (a) shown in the extended 
Brillouin zone $(p_y, p_z)$ for $N=1$ and 
$\omega^*_c(\theta , \phi) / \omega_b(\theta) = 15$ 
[see Eqs. (3),(7)], the velocity component 
$v_z(t)$ (b) is a periodic 
function of $t \sim p_y$ with an average value 
$ < v_z(t) >_t$ being given by Eq.(9). 
Electron transport along ${\bf z}$-axis, $z(t)$, (c)
is a step-like function of $t \sim p_y$ and is defined by the interference 
effects between "effective stripes" (ES) located near the following two
 series of points in the extended Brillouin zone:
$p_y = acrsin[\omega_c(\theta,\phi)/\omega^*_c(\theta,\phi)] + 2 \pi n$
and $p_y = \pi - acrsin[\omega_c(\theta,\phi)/\omega^*_c(\theta,\phi)] 
+ 2 \pi n$, where $n$ is an integer.
As it follows from Fig.1c, the ES are "in phase" in the extended 
Brillouin zone $(p_y , p_z)$ for the "commensurate" electron orbits given
by Eq.(3).}
\label{fig1}
\end{figure}

\begin{figure}[h]
\includegraphics[width=2.8in,clip]{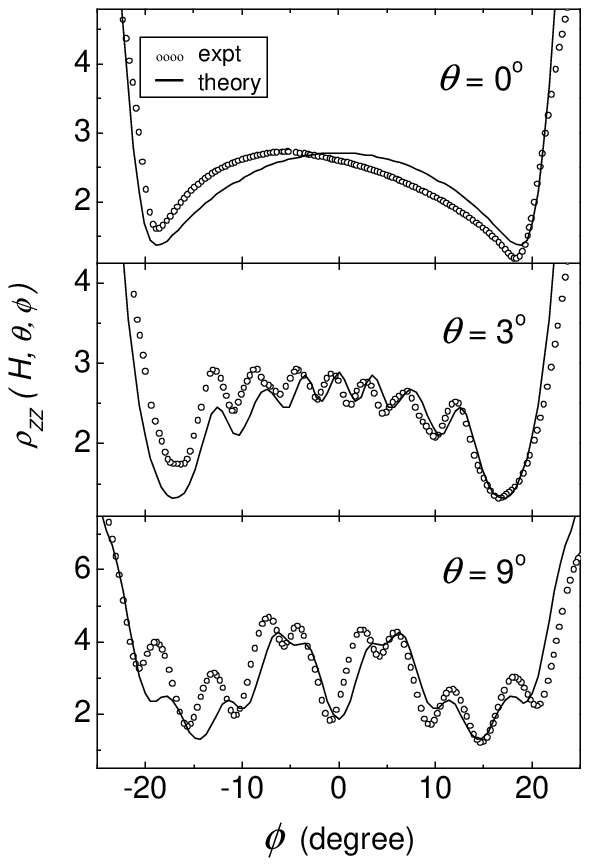}
\caption{ IC oscillations of resistivity $\rho_{zz}(\theta,\phi)$ calculated for 
(a) $\theta = 0^o$, (b) $\theta = 3^o$, and (c) $\theta = 9^o$ by means of Eqs. (17),(18) (solid lines) are compared with the experimental data [15] (dotted lines).
Note that at $\theta = 9^o$ $\rho_{zz}$ minima occur at
all integers $N$ in Eq.(3), whereas at $\theta = 3^o$ they
correspond only to odd integers $N$ in Eq.(3). }
\label{fig1}
\end{figure}


\begin{references}

\bibitem{IY} T. Ishiguro, K. Yamaji, and G. Saito, 
{\it Organic Superconductors} 
(Second Edition, Springer-Verlag, Heidelberg, 1998).

\bibitem{Shc}See review articles in 
{\it Common Trends in Synthetic Metals and High-Tc 
Superconductors} 
[I. F. Schegolev's Memorial volume of J. Phys. I (France) 
\underline{6} (1996)] and references therein.

\bibitem{Br} See recent review S.E. Brown, M.J. Naughton, I.J. Lee, 
E.I. Chashechkina, and P.M. Chaikin
in {\it More is Different} (Fifty Years of Condensed Matter
Physics), N.P.  Ong and R.N. Bhatt Eds.
(Princeton Series in Physics, Princeton, 2001) 
and references therein.

\bibitem{Mag1} A.G. Lebed, Pis'ma Zh. Eksp. Teor. Fiz.
\underline{43}, 137 (1986) [JETP Lett. \underline{43}, 174 (1986)];
A.G. Lebed and Per Bak, Phys. Rev. Lett. 
\underline{63}, 1315 (1989).

\bibitem{Lbak} A.G. Lebed, J. Phys. I (France) 
\underline{4}, 351 (1994);
J. Phys. I (France) \underline{6}, 1819 (1996).


\bibitem{Mike} M.J. Naughton, O.H. Chung, L.Y. Chiang,
and J.S. Brooks,
Mat. Res. Soc. Symp. Proc. \underline{173}, 257 (1990).

\bibitem{Boe} G.S. Boebinger, G. Montambaux, M.L. Kaplan,
R.C. Haddon, S.V. Chichester, and L.Y. Chiang,
 Phys. Rev. Lett. \underline{64} 591 (1990).

\bibitem{Osa} T. Osada, A. Kawasumi, S. Kagoshima, N. Miura,
and G. Saito, 
Phys. Rev. Lett. \underline{66} 1512 (1991).

\bibitem{} See recent papers 
E.I. Chashechkina and P.M. Chaikin,
Phys. Rev. Lett. \underline{80}, 2181 (1998);
Phys. Rev. B \underline{65}, 012405 (2002) and references
therein.

\bibitem{DKC} G.M. Danner, W. Kang, and P.M. Chaikin,  
Phys. Rev. Lett.  \underline{72} , 3714 (1994).

\bibitem{ZM} T. Osada, S. Kagoshima, and  N. Miura, 
Phys. Rev. Lett. \underline{77} , 5261 (1996). 

\bibitem{AK} M.J. Naughton, I.J. Lee, P.M. Chaikin, 
and G.M. Danner, 
Synth. Met. \underline{85} , 1481 (1997).


\bibitem{Osa} H. Yoshino, K. Saito, K. Kakuchi, 
H. Nishikawa, K. Kobayashi, and I. Ikemoto, 
J. Phys. Soc. Jpn \underline{64} , 2307 (1995).

\bibitem{LB} A.G. Lebed and N.N. Bagmet, 
Phys. Rev. B \underline{55}, R8654 (1997).


\bibitem{K} I.J. Lee and M.J. Naughton, Phys. Rev. B 
\underline{57}, 7423 (1998). 

\bibitem{AK} I.J. Lee and M.J. Naughton, Phys. Rev. B 
\underline{58}, R13343 (1998).


\bibitem{Y1} H. Yoshino, K. Murata, T. Sasaki, K. Saito,
H. Nishikawa, K. Kikuchi, K. Kobayashi, and Isao Ikemoto,
J. Phys. Soc. Jpn \underline{66}, 2248 (1997).


\bibitem{Os} T. Osada, N. Kami, R. Kondo, and S. Kagoshima,
Synth. Met. \underline{103}, 22024 (1999).


\bibitem{ZM} H. Yoshino and K. Murata, 
J. Phys. Soc. Jpn \underline{68}, 3027 (1999). 

\bibitem{DC} G.M. Danner and P.M. Chaikin, Phys. Rev. Lett. 
\underline{75}, 4690 (1995). 

\bibitem{SCA} S.P. Strong, D.G. Clarke, and P.W. Anderson,
 Phys. Rev. Lett. \underline{73}, 1007 (1994). 

\bibitem{DOC} G.M. Danner, N.P. Ong, and P.M. Chaikin, 
Phys. Rev. Lett. \underline{78}, 983 (1997). 

\bibitem{Ab} See, for example, A.A. Abrikosov, 
{\it Fundamentals of Theory of Metals} (Elsevier Science
Publisher B.V., Amsterdam, 1988).

\bibitem{OKM} T. Osada, S. Kagoshima, and N. Miura, Phys. Rev. B 
\underline{46}, 1812 (1992). 


\bibitem{MVK} M.V. Kartsovnik et al., J. Phys. I (France) 
\underline{2}, 223 (1992). 

\bibitem{R1} We don't take into account a $p_y$-dependence 
of the velocity component $v_x(p_y)$ as well as omit the Lorentz-force 
component resulting from the velocity component $v_z(p_z)$. 
Strict analysis shows that the former approximation is valid for small enough values of angle, $\phi \leq 20^o$, with an accuracy $ \leq 10^{-2} $.
The latter approximation is equivalent to disregarding small closed orbits 
in an inclined magnetic field (2) [14]. 

\bibitem{R3} In contrast to the MA orbits [4], "commensurate" 
trajectories (3),(7) are periodic only in a 
reciprocal plane $(p_y , p_z)$ and do not correspond to a periodic 
motion in a real $(y,z)$-plane.
 

\bibitem{GL} L.P. Gor'kov and A.G. Lebed, 
J. Phys. Lett. (France) \underline{45}, L-433 (1984).

\bibitem{MA} For quasi-classical theories of the MA effects [4], 
see Refs. [30,24,18].
 
\bibitem{LN} A.G. Lebed, in preparation.

\end{references}
\end{document}